\documentclass[prl,showpacs,twocolumn,superscriptaddress,floatfix,amsmath]{revtex4}
\usepackage[latin1]{inputenc}
\usepackage[english]{babel}
\usepackage{graphicx,psfrag}
\usepackage{amsmath}
\usepackage{amssymb}
\usepackage{amscd}
\usepackage{eucal}
\usepackage{color}
\usepackage{bm}

\begin{document}
\title{Electrical probing of the spin conductance of mesoscopic cavities}
\author{\.{I}. Adagideli}
\affiliation{Institut f\"ur Theoretische Physik, Universit\"at Regensburg,
D-93040, Germany}
\author{J.~H.~Bardarson}
\affiliation{Instituut-Lorentz, Universiteit Leiden, P.O. Box 9506,
2300 RA Leiden, The Netherlands}
\author{Ph.~Jacquod}
\affiliation{Physics Department,
   University of Arizona, 1118 E. 4$^{\rm th}$ Street, Tucson, AZ 85721, USA}
\date{\today}
\begin{abstract}
We investigate spin-dependent transport in multiterminal mesoscopic cavities
with spin--orbit coupling. Focusing on
a three-terminal setup we
show how injecting a pure spin current or a polarized current from one
terminal generates additional charge current and/or voltage across the two
output terminals.
When the injected current is a pure spin current, a single measurement
allows to extract the spin conductance of the cavity.
The situation is more complicated for a polarized injected current, and we show
in this case how two purely electrical measurements on the output currents,
give the amount of current that is solely due to spin-orbit interaction. This
allows to extract the spin conductance of the device also in this case.
We use random
matrix theory to show that the spin conductance of chaotic ballistic cavities
fluctuates universally about zero mesoscopic average and describe
experimental implementations of mesoscopic spin to charge current converters.
\end{abstract}
\maketitle{}

Many recent theoretical, experimental and numerical investigations have
explored possibilities to generate spin currents and accumulations
in spin-orbit coupled
diffusive~\cite{Dya71,Sin04,Mur04,Ino04,Mis04,Ada05,Kat04,Wun05,Sai06,Sch05,Rai05,Ren06,Val06,Ada07}
and ballistic~\cite{Bar07,Nik05,Naz07,Kri08} systems.
The main focus of this field of {\it spin-orbitronics}
is on purely electrostatic generation of
spin currents via application of charge currents
and/or voltage biases at appropriate lead contacts to the device.
The amount of spin current generated by a given bias defines
a {\it spin conductance} $G^{(s)}$ characterizing the spin generation efficiency
of
the device. Although $G^{(s)}$ is theoretically convenient, no
realistic setup to experimentally probe it has been proposed to date.
Such a setup is highly desirable for ballistic mesoscopic cavities,
which typically feature  relatively high spin conductances~\cite{Bar07,Kri08}.
This
property is however
not sufficient to make them
good candidate components for low-power spintronic devices because of the
relatively large mesoscopic
fluctuations exhibited by their spin
conductance~\cite{Bar07}. Originating from the phase coherence of spin
transport,
these fluctuations are beyond the existing measurement
proposals which are based on theories describing ensemble averaged
diffusive transport of spins, assuming locally
well defined spin accumulations~\cite{Han04,Erl05,Ada06}.
In this Letter, we propose to use spin-orbit coupled ballistic
mesoscopic cavities as spin- to charge-current converters to
experimentally analyze spin currents and spin accumulations in
meso- and nanoscopic devices. We show how the
spin conductance of such cavities can be directly measured
from the amount of charge current they generate out of
conventionally injected spin currents.

\begin{figure}
\includegraphics[width=\columnwidth]{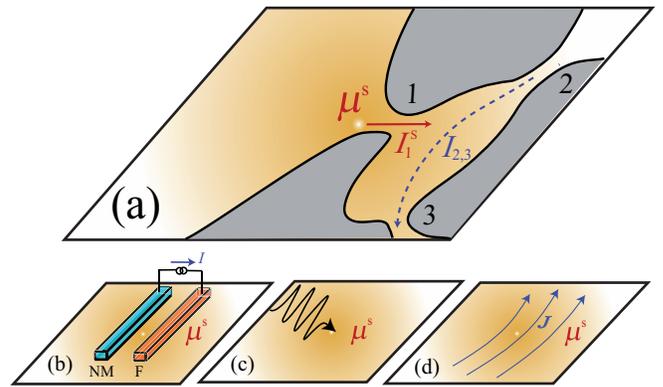}
\caption{\label{fig1} (Color online) (a) Three terminal ballistic quantum dot as
a
mesoscopic spin current to charge current converter.
The output electrodes 2 and 3 are at the same fixed
potential $V=0$, while the input electrode 1 is at potential $V_1$.
Spins are injected into the dot from
electrode 1. We argue that spin-orbit coupling can convert this spin current
into a charge current across electrodes 2 and 3 (dashed blue arrow).
We discuss three possible mechanisms for generating this spin current
from a spin accumulation generated by (b) a ferromagnet,
(c) polarized photons and (d) a charge current in the spin-orbit coupled
reservoir 1. For these three mechanisms, it is straightforward to invert
the spin accumulation, allowing to extract the spin conductance of the cavity
from two measurements of the output charge current/voltage
at opposite spin accumulations.}
\end{figure}

For simplicity, we choose the spin- to charge-current
converter to be an open three-terminal quantum
dot with spin-orbit coupling -- this is sketched in Fig.~\ref{fig1}a -- though
our discussion is straightforwardly generalized to
multi-terminal cavities with any number of leads greater than two.
The spin accumulation
$\boldsymbol{\mu}^{(s)}$
(the components of that vector give the difference in
chemical potential of different spin species along the corresponding spin direction)
in a bulk electron reservoir generates a
spin current $I_1^{(s)}$ injected into the dot from terminal 1~\cite{Ada07}.
Spin-orbit coupling inside the
dot acts on the pure spin part of this current
in a manner similar to the inverse spin Hall effect~\cite{Val06} --
it converts
it into either a transverse charge current,
or a voltage difference between lead 2 and lead 3.
This conversion differs however
from the inverse spin Hall effect in that it is fully coherent and it couples different spin
polarization.
Measuring this charge current/voltage allows to extract the spin
conductance of the cavity.
We consider two types of measurements, defining two different spin conductances
$G^{(s1)}$ and $G^{(s2)}$.
In the first one, the voltage
on terminal 1 is set such that $I_1^{(s)}$ is a pure spin current,
without charge component, $I_1=0$. Then
$I_2 = - I_3$ are entirely
due to the conversion of $I_1^{(s)}$ into a charge current, and the spin
conductance of the cavity is defined as
$G^{(s2)}_j \equiv I_j/|\boldsymbol{\mu}^{(s)}|$, $j=2,3$.
In the second scenario,
$I_1^{(s)}$ is a polarized current, accompanied by a net injection of
charges into the dot, $I_1 \ne 0$. To demonstrate the existence of a spin
component in $I_1^{(s)}$, one thus needs to isolate that part of $I_2$ or $I_3$
that originates exclusively from the spin-orbit conversion of $I_1^{(s)}$.
This is achieved by performing two measurements at reversed
spin accumulation in 1, $\boldsymbol{\mu}^{(s)} \rightarrow -\boldsymbol{\mu}^{(s)}$,
but fixed electrochemical potentials. As we show below,
the difference in the two measured currents is
solely due to the spin-orbit conversion of $I^{(s)}$. This
defines the spin conductance of the cavity as
$G_j^{(s1)} \equiv
[I_j(\boldsymbol{\mu}^{(s)})-I_j(-\boldsymbol{\mu}^{(s)})]/
2|\boldsymbol{\mu}^{(s)}|$, $j=2,3$,
where the upper index refers
to the type of measurement, while the lower index refers to the
exit lead on which the current measurement is performed.
In both instances,
we show that $G^{(s)}_j$ exhibits mesoscopic fluctuations
about zero average, under variation of the shape of the cavity or
homogeneous changes in electrochemical potentials in all terminals.

There are various ways to generate spin accumulations and currents, most
notably via spin injection from ferromagnetic
components~\cite{Lou06} or optical orientation~\cite{optical},
or via magneto-electric effects~\cite{Ede90,Aro89,Duc08,Kat05}.
Fig.~\ref{fig1}b illustrate the first method, where
a ferromagnetic lead, F, and a nonmagnetic lead, NM,
form an $F\big/2DEG(2DHG)\big/NM$ junction with a two dimensional
electron (hole)
reservoir. Passing a current
between F and NM injects spins into the reservoir.
In Fig.~\ref{fig1}c we sketch how spin polarization is generated via
optical pumping with circularly polarized photons. Fig.~\ref{fig1}d
illustrates how a steady-state electronic current flowing in a
two-dimensional $k$-linear spin-orbit coupled system generates a
bulk spin accumulation~\cite{Ede90}.
In all cases, the spin accumulation, generated a distance shorter than the spin
relaxation
length but longer than the mean free path away from the point contact,
diffuses to and flows through the ballistic cavity.
Then, the ballistic processes connecting the spin injector part of the circuit
with the
cavity can be ignored and the reservoir can be viewed as having a well-defined
spin
accumulation $\boldsymbol{\mu}^{(s)}$.

We formalize our theory. An open quantum dot is coupled to three bulk reservoirs
via
ideal point contacts, each carrying $N_i$ open channels ($i=1,2,3$).
We assume that spin-orbit coupling exists only inside the dot. Given that
the dot and the reservoirs are made of the same material, this is justified
when (i) the openings to the electrodes are small enough, that the spin-orbit
time is shorter than the mean dwell time spent by an electron in the dot, and (ii)
the accumulations in the reservoirs are generated a distant shorter than the
spin-orbit length away from the dot.
We follow the scattering approach to transport and start from
the linear relation between currents and
chemical potentials~\cite{But86},
\begin{equation}\label{scatt0}
I_i^{(\alpha)} = \frac{e}{h} \sum_{\beta}(2N_i \delta_{\alpha\beta}-\mathcal{T}^{(\alpha\beta)}_{ii})\mu_{i}^{(\beta)}
-\frac{e}{h} \sum_{j \ne i,\beta} \mathcal{T}^{(\alpha\beta)}_{ij}
\mu_{j}^{(\beta)}.
\end{equation}
Here, $\mu_i^{(0)} = e V_i$ ($V_i$ is the voltage applied to terminal $i$) and
$\mu_i^{(\beta)}$ are the components of
the spin
accumulation vector $\boldsymbol{\mu}^{(s)}_{i}$,
giving the difference in chemical potential between the two spin species
along the corresponding axis, i.e. $\mu_i^{(z)} = \mu_i^{(\uparrow)}
- \mu_i^{(\downarrow)}$,
while $I^{(0)}_i \equiv I_i$ and $I_i^{(\alpha)}$ are the charge current
and the components
of the spin current vector, all evaluated in terminal $i$.
We introduced the spin-dependent transmission probabilities
\begin{equation}
\mathcal{T}_{ij}^{(\alpha\beta)} =\sum_{m \in i,n \in j}
{\rm Tr} [ t_{mn}^\dagger
 \sigma^{(\alpha)}_m t_{mn} \sigma^{(\beta)}_n],
\end{equation}
where $\sigma^{(\alpha)}$, $\alpha = 0,x,y,z$ are Pauli matrices
($\sigma^{(0)}$ is the identity matrix)
whose lower index indicates whether they apply to the spin
components at the entrance or exit lead,
the trace is taken over the spin degree of freedom and $t_{mn}$ is a 2$\times$2
matrix of spin-dependent transmission amplitudes from channel $n$ in
lead $j$ to channel $m$ in lead $i$. In Ref.~\cite{Bar07},
only transmission probabilities $\mathcal{T}_{ij}^{(\alpha0)}$ were considered,
because the reservoirs had no spin accumulation,
and consequently spin currents were determined by a single polarization direction.

We need to determine the chemical potentials.
First, reservoir 1 is kept at a fixed voltage
$V_1$ and
spin accumulation $\boldsymbol{\mu}_1^{(s)} \equiv
\boldsymbol{\mu}^{(s)}$.
Second,
we set the electrochemical potentials to zero in reservoirs 2 and 3.
Third, because the leads are ideally connected
to the dot, and because reservoirs 2 and 3 see
no source of spins other than the one injected from the cavity, we also set
the spin accumulations of the reservoirs $2$ and $3$ to zero.
Under these conditions, the components of $I_1^{(s)}$ are
\begin{eqnarray}
I_1^{(\alpha)} & = &  \frac{e}{h} \sum_{\beta}(2N_1 \delta_{\alpha\beta}-\mathcal{T}^{(\alpha\beta)}_{11})\mu_{1}^{(\beta)}.
\end{eqnarray}
Unless very specific conditions are met, $I_1^{(s)}$ is finite.
The charge currents in lead $j=2,3$ are
\begin{equation}
I_j=-\frac{e^2}{h} \mathcal{T}^{(00)}_{j1} V_{1}-\frac{e}{h}\sum_{\beta\ne0}
\mathcal{T}^{(0\beta)}_{j1}\mu_{1}^{(\beta)}.
\end{equation}
The first contribution to $I_j$ is the well known nonlocal charge
conductance of the cavity. We are mostly interested in the
second contribution which corresponds to the conversion of the spin
accumulation to charge current. In order to
extract the spin conductance of the cavity from the current measurement, we
isolate this second contribution by switching the polarization direction
$\boldsymbol{m}$
of the spin accumulation.

This can be achived e.g. for ferromagnetic injection by temporarily
applying an external magnetic field in the appropriate direction.
Within linear response in the magnetic field,
the only
effect of doing this is to switch the direction of the spin accumulation
in reservoir 1, $\boldsymbol{\mu}^{(s)} = \mu^{(s)} \boldsymbol{m}
\rightarrow-\boldsymbol{\mu}^{(s)} $,
without changing its voltage bias, $V_1 \rightarrow V_1$.
The spin conductance of the cavity,
\begin{equation}\label{cond_def1}
G^{(s1)}_j=\big(I_j(\boldsymbol{m})-I_j(-\boldsymbol{m})\big)/2\mu^{(s)}=-\frac{e}{h}\sum_{\beta\ne0}
\mathcal{T}^{(0\beta)}_{j1} m^{(\beta)}
\end{equation}
is then directly extracted from the difference in the charge current in
lead $j=2,3$ between these two measurements.
This first definition of the
spin conductance is appropriate only when
the effect of the applied magnetic field can be treated
in linear response.

When the spin injection part of the circuit is not operating within
linear response,
inverting the magnetization direction results in different magnitudes of
the chemical  potentials. In this regime we instead apply a charge voltage
bias on lead 1 such that the current through it vanishes, $I_1 = 0$.
Then the pure spin current that flows through lead 1
generates a charge current flowing from lead 2 to lead 3 giving
a spin conductance
\begin{equation}\label{cond_def2}
G^{(s2)}_j =-\frac{e}{h}\sum_{\beta\ne0}
\left(\frac{\mathcal{T}^{(00)}_{j1} \; \mathcal{T}^{(0\beta)}_{11}}{2N_1-
\mathcal{T}^{(00)}_{11}} +\mathcal{T}^{(0\beta)}_{j1}
\right)m^{(\beta)}.
\end{equation}
It is remarkable that when $N_1=1$, both definitions of the spin conductance
are equal and one has
$G^{(s1)}_2 = G^{(s2)}_2
= -G^{(s1)}_3 = - G^{(s2)}_3$, because then time-reversal symmetry
imposes $\mathcal{T}^{(0\beta)}_{11}=0$,
$\forall \beta \ne 0$. Eqs.~(\ref{cond_def1},\ref{cond_def2}) are general and do not
rely on any
assumption on the charge/spin dynamics in the cavity.

From now on we focus on the experimentally relevant case of
a coherent quantum dot with chaotic ballistic electron dynamics.
Accordingly, we use
random matrix theory (RMT) to calculate the average and
fluctuations of $I_{2,3}$~\cite{Bro96}.
RMT replaces the system's scattering matrix $S$ -- whose elements are
given by the transmission amplitudes $t_{mn}$, as well as
reflection amplitudes --
by a random unitary matrix. Our interest resides on systems with
time reversal symmetry (absence of magnetic field)
and totally broken spin rotational symmetry (strong spin-orbit coupling),
as in the experiments of Refs.~\cite{Marcus,Grbic}.
In this case $S$ is an element of
the circular symplectic ensemble
(CSE).  Following Ref.~\cite{Bar07},
we rewrite the generalized transmission
probabilities $\mathcal{T}_{ij}^{(\alpha \beta)}$
as a trace over $S$
\begin{align}
  \label{eq:TfullTr}
  \mathcal{T}_{ij}^{(\alpha \beta)} &=
\text{Tr}\,[Q_i^{(\alpha)}SQ_j^{(\beta)}S^\dagger], \\
  [Q_i^{(\alpha)}]_{m\mu,n\nu} &= \notag
  \begin{cases}
    \delta_{mn}~\sigma^{(\alpha)}_{\mu \nu}, & \sum_{j=1}^{i-1}N_j < m \leq
\sum_{j=1}^{i} N_j, \\
    0, & \text{otherwise}.
  \end{cases}
\end{align}
Here, $m$ and $n$ are channel indices, while
$\mu$  and $\nu$ are spin indices. The trace is taken
over both sets of indices.

Averages, variances, and covariances of the generalized transmission
probabilities, Eq.~\eqref{eq:TfullTr}, over the CSE can be calculated using the
method of Ref.~\cite{Bro96}. Experimentally, these quantities correspond to
an ensemble of measurements on differently shaped quantum dots at different
global electrochemical potentials. The RMT--averaged transmission
probabilities read
\begin{equation}
  \label{eq:Tmean}
  \langle \mathcal{T}^{(\alpha \beta)}_{ij} \rangle =
\frac{2 \delta_{\alpha\beta}}{N_T-1/2}\left[N_iN_j
\delta_{\alpha 0} - (\delta_{\alpha 0}-1/2)N_i\delta_{ij}\right].
\end{equation}
Together with the covariances
$\langle \delta \mathcal{T}_{ij}^{(00)} \delta \mathcal{T}_{kl}^{(0\beta)}
\rangle \propto \delta_{\beta0}$, we readily obtain that the
spin conductances vanish on average,
$\langle G^{(s)}_{i}(j) \rangle=0$.
They nevertheless fluctuate from sample to sample or upon global
homogeneous variation of the electrochemical potentials, and we thus
calculate $\text{var}\;G^{(s)}_{i}(j) $. For $\beta \ne 0$ one gets
($N_T=N_1+N_2+N_3$)
\begin{subequations}
\begin{eqnarray}
&&{\rm var} \; \mathcal{T}_{11}^{(0\beta)}=\frac{4 N_1 (N_1-1) (N_T-N_1)}{N_T (2
N_T-1) (2 N_T-3)}, \\
&&{\rm var} \; \mathcal{T}_{j1}^{(0\beta)}=\frac{4 N_j N_1 (N_T-N_1-1)}{N_T (2
N_T-1) (2 N_T-3)}, \, (j\ne 1), \\
&&\left\langle \frac{{\cal T}_{j1}^{(00)} }{2 N_1 - {\cal T}_{11}^{(00)}}
\right \rangle
\simeq \frac{N_j}{N_T-1},
\end{eqnarray}
\end{subequations}
from which we obtain
\begin{subequations}
  \label{varG}
\begin{eqnarray}
  \label{varG1}
  \text{var}\;G^{(s)}_{1}(j) &=&\frac{e^2}{h^2}
\frac{4 N_j N_1 (N_T-N_1-1)}{N_T (2 N_T-1) (2 N_T-3)}, \\
\label{varG2}
  \text{var}\;G^{(s)}_{2}(j) & \simeq &\text{var}\;G^{(s)}_{1}(j) \\
&+& \frac{e^2}{h^2}
\frac{4 N_j^2 N_1 (N_1-1)}{N_T (2 N_T-1) (2 N_T-3)(N_T-1)}.\nonumber
\end{eqnarray}
\end{subequations}
To obtain Eq.~(\ref{varG2}), we once again noted that
$\langle \delta \mathcal{T}_{ij}^{(00)} \delta \mathcal{T}_{kl}^{(0\beta)}
\rangle \propto \delta_{\beta0}$, and neglected the
subdominant fluctuations of
${\cal T}_{j1}^{(00)}\big/(2 N_1 - {\cal T}_{11}^{(00)})$. The second term
in this equation is a leading order approximation in $N_i^{-1}$.
One can show however that it is always
significantly smaller than ${\rm var} \, G_1^{(s)}(j)$, for any $N_i$, so that
the small deviations
from Eq.~(\ref{varG2}) possibly occurring for small number of channels do not
alter our conclusions.
We see that, while the conductance
across electrodes 2 and 3 vanishes on
RMT average,
it exhibits sample-to-sample fluctuations. These fluctuations are universal in
the common
mesoscopic sense that they remain the same if the number of channels carried by
all leads
is homogeneously rescaled. For a given sample, the
conductance is thus finite, and can be approximated by
its typical value $ \approx {\rm rms} \;
G^{(s)}_{i}(j)$, $i=1,2$.
In this article we used the definitions of
Eqs.~(\ref{cond_def1})
and (\ref{cond_def2}) that spin conductances are
given by the ratio of a charge current with
a spin accumulation.
Converted into more standardly used units of
spin conductance,
and for a symmetric cavity with $N_i = N \gg 1$, $\forall i$,
Eq.(\ref{varG1}) predicts a typical spin conductance of
${\rm rms} \, G_1^{(s)} \simeq \sqrt{2/27} (e/h) \rightarrow \sqrt{2/27}
(e/4 \pi)$.

Instead of measuring $I_{2,3}$ for $V_{2,3}=0$, one can alternatively
tune $V_{2,3}$ such that the currents vanish. Going back to Eq.~(\ref{scatt0}),
one obtains that the potential difference $\delta V_2 \equiv V_2({\bf m}) -
V_2(-{\bf m})$
satisfies
\begin{eqnarray}\label{I0}
e \delta V_2 &=&
2 \sum_{\beta \ne 0} \mathcal{T}^{(0\beta)}_{31} \mu_{1}^{(\beta)} \\
&\times & \left[2N_2 -\mathcal{T}^{(00)}_{22}-\mathcal{T}^{(00)}_{23} \,
\frac{2N_2 -\mathcal{T}^{(00)}_{22}+\mathcal{T}^{(00)}_{32}}
{2N_3 -\mathcal{T}^{(00)}_{33}+\mathcal{T}^{(00)}_{23}} \right]^{-1}. \nonumber
\end{eqnarray}
For small number of channels per lead, say $N_i \lesssim 5$, Eq.~(\ref{I0})
gives a
voltage response similar in magnitude to the spin accumulation in reservoir 1.
In 2DEG/2DHG, a
response in the range $\sim$0.1--1 $mV$ thus
typically requires a spin accumulation
of the order 0.001--0.1 $E_{\rm F}$, with 0.1-1\% of
polarized electrons, depending on the material.
This is certainly achievable via optical pumping,
where polarization of significant fractions of the
electronic gas has been demonstrated~\cite{Sti07}, and is reasonably expectable
for ferromagnetic injection, based on polarizations obtained in bulk
semiconductors~\cite{Lou06} (though ferromagnetic injection into a
2DEG/2DHG has yet to be demonstrated).
Experimental measurements of the Rashba parameter in InAs-based
2DEG~\cite{Nitta} give a ratio of the spin-orbit splitting energy to
Fermi energy of the order of 5 meV/100meV = 1/20. Given a carrier
concentration of $n_s = 2 \cdot  10^{12} {\rm cm}^{-2}$,
we estimate that the Edelstein mechanism~\cite{Ede90} would produce
polarizations of the order of 0.1-1 \% in a 0.2 $\mu m$ wide strip
of InAs-based 2DEG carrying a current of about 2000 $n A$.
Therefore the spin-to-charge current conversion discussed in this
article leads to measurable charge voltage differences.

In conclusion, we have discussed how spin currents or spin accumulations
can be converted
mesoscopically to charge currents and voltages using neither ferromagnets
nor external magnetic fields. We have proposed an experimental method,
based on this spin-charge conversion, to measure the spin conductance of
mesoscopic cavities
-- giving the charge current generated solely
by the presence of spin-orbit interaction --
relying solely on measuring electrical signals.
The spin conductance of a mesoscopic cavity might in principle be measured
using spin polarized quantum point contacts~\cite{Koo08}. However,
the Zeemann field necessary to polarize the quantum point contact
is rather large. It might thus
freeze the spin of the electrons and reduce or even destroy
spin-orbit effects
inside the cavity. The setup we propose does not suffer from this.
We do not see any unsurmountable difficulty preventing the experimental
implementation of the ideas presented here.

This work has been partially supported by the National Science
Foundation under Grant No. DMR--0706319 and by the German science foundation DFG
under grants SFB689 and GRK638. We would like to thank R.~Leturcq
who gave us the initial motivation to work out
these ideas and D. Weiss for discussions.
PJ thanks the Physics Department of the
Universities of Basel and Regensburg for their kind hospitality in the final
stages of this
project.

\end{document}